\documentclass[%
 reprint,
 amsmath,amssymb,
 aps,
]{revtex4-2}

\usepackage{graphicx}
\usepackage{dcolumn}
\usepackage{bm}

\begin{document}

\preprint{APS/123-QED}

\title{Exciton--Exciton and Exciton--Photon Annihilation in Polaritonic Systems} 

\author{Luca Nils Philipp}
\affiliation{Institut für Physikalische und Theoretische Chemie, Universität Würzburg, Emil-Fischer Straße 42, 97074 Würzburg, Germany}%
\author{Julian Lüttig}
\affiliation{Department of Physics, University of Ottawa, Ottawa, ON K1N 6N5, Canada}%
\author{Roland Mitri\'c}
\affiliation{Institut für Physikalische und Theoretische Chemie, Universität Würzburg, Emil-Fischer Straße , 97074 Würzburg, Germany}%

\date{\today}

\begin{abstract}
Strong light--matter interactions forming hybrid quasiparticles termed polaritons can specifically tailor molecular photophysics. In this spirit, enhancing energy transport has recently been of special interest. Exciton--exciton annihilation is commonly used to quantify energy transfer in excitonic systems, and has been recently applied to investigate transport dynamics in polaritonic systems. However, the interpretation of experimental findings is challenging without a microscopic theory that accounts for the various nonradiative relaxation channels determining the quasiparticle diffusion length. In this work, we develop a microscopic model for polariton annihilation based on exciton--exciton annihilation and propose an exciton--photon annihilation as the decisive process that competes with exciton--exciton annihilation. The interplay between exciton--exciton and exciton--photon annihilation ultimately governs the annihilation dynamics and sets the fundamental limit to the transport efficiency. Our model explains recent experimental results and demonstrates that increased annihilation rates might serve as an explicit fingerprint to differentiate between the response of polaritons and other untargeted effects.
\end{abstract}

\maketitle
Energy transfer is a fundamental process enabling functionality of various systems such as solar cells \cite{forrest_path_2004, menke_tailored_2013}, natural photosynthetic systems \cite{scholes_lessons_2011, mirkovic_light_2017} or artificial light-harvesting materials \cite{brixner_exciton_2017, scholes_excitons_2006}. The coupling of quantized light modes with delocalized molecular excitations (excitons) has proven to be a successful strategy to alter and control the properties of the emergent quasiparticles \cite{thomas_ground-state_2016, ahn_modification_2023, galego_cavity-induced_2015, hutchison_modifying_2012, schwartz_reversible_2011, feist_extraordinary_2015, schachenmayer_cavity-enhanced_2015, zhong_non-radiative_2016, nagarajan_conductivity_2020, orgiu_conductivity_2015, rozenman_long-range_2018, berghuis_controlling_2022}. These quasiparticles arise by the coupling of electronically excited molecular states to the field states of a cavity and are commonly referred to as molecular polaritons \cite{fassioli_femtosecond_2021}. It has been theoretically and experimentally proposed that the formation of molecular polaritons can enhance energy transport \cite{feist_extraordinary_2015, schachenmayer_cavity-enhanced_2015, berghuis_controlling_2022, wu_efficient_2025, zhong_energy_2017, aroeira_theoretical_2023}. Probing ultrafast transport involving polaritons is challenging, since the transport does not necessarily result in changes in the absorption spectrum, and therefore standard techniques such as transient absorption spectroscopy may not be suited to probe energy transfer. To solve this problem a variety of different techniques have been developed such as spatially resolved transient absorption spectroscopy \cite{schnedermann_ultrafast_2019} or spectrally resolved fluorescence quenching \cite{lunt_exciton_2009}. Another successful strategy commonly used in excitonic systems is probing the transfer indirectly via exciton--exciton annihilation \cite{scheblykin_non-coherent_1998, dostal_direct_2018, kriete_interplay_2019, rehhagen_exciton_2020}. This process involves two excitons in which the energy of one exciton is dissipated nonradiatively into heat \cite{may_kinetic_2014}. Since the interaction of two excitons can only occur if  they are in close proximity to each other, annihilation can be used as a probe to monitor the exciton diffusion. A similar approach has been applied to polaritonic systems investigating enhanced transfer processes between photosynthetic light-harvesting (LH2) complexes strongly coupled to a Fabry--Pérot cavity \cite{wu_efficient_2025}. However, polaritonic systems exhibit a complex electronic structure and lately it has been questioned whether inherent features of polaritons or some other untargeted effects such as the response of so-called dark states are experimentally observed \cite{chen_tracking_2025, renken_untargeted_2021}. Thus, a major current question of polariton photophysics is to identify characteristic properties of polaritons that can be utilized as their experimental fingerprint. The multi-particle dynamics of polaritons might be such a characteristic property. Thus, a microscopic theoretical description of polariton annihilation is highly desirable. Furthermore, a theoretical description is crucial to understand how experimental signatures connect to the microscopic energy transfer. Modeling polariton annihilation allows one to understand how transfer, and thus annihilation, is altered by the system parameters such as the coupling to the light mode. While for exciton--exciton annihilation a well established theoretical framework has been developed \cite{ryzhov_low-temperature_2001, tempelaar_excitonexciton_2017, scheblykin_non-coherent_1998, may_kinetic_2014, kumar_exciton_2023}, a detailed microscopic description of polariton annihilation is missing so far.

\begin{figure}[t]
  \includegraphics[]{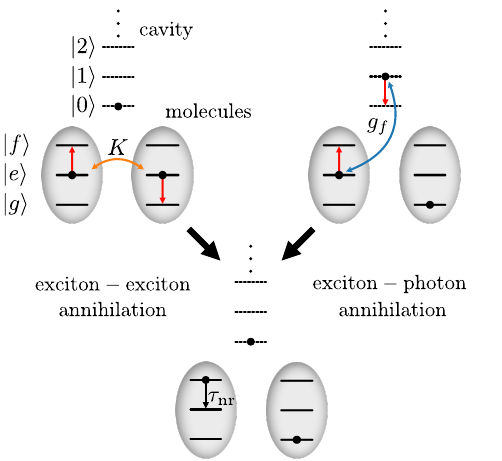}
  \caption{\label{figure1}Microscopic representations of exciton--exciton and exciton--photon annihilation in polaritonic systems. Either two singly excited molecules couple resonantly through Coulomb interactions with coupling strength $K$ or a singly excited molecule and a cavity photon couple resonantly through light--matter interactions with coupling strength $g_f$ such that a highly excited molecular state is formed by both processes. Subsequently, the resulting excited molecular state undergoes fast internal conversion with rate $1/\tau_\mathrm{nr}$ into the first excited molecular state.}
\end{figure}

In this letter, we provide a theoretical model to describe the annihilation of polaritons and investigate its dependence on various system parameters. We use the well-known theory of exciton--exciton annihilation as a starting point. To this end, consider an ensemble of molecules, where each molecules is represented by three states, i.e., a ground state $|g\rangle$ together with a first excited $|e\rangle$ and a second excited $|f\rangle$ state. The process of exciton--exciton annihilation can be dissected into different steps. First, two excitations need to be in close proximity in such a way that interaction between them can occur. In a second step, exciton--exciton interaction leads to the fusion of the excitons resulting in the formation of the higher excited molecular state $|f\rangle$. The annihilation is ultimately enabled by a fast internal conversion from the $|f\rangle$ state to the first excited molecular state $|e\rangle$. This process is schematically depicted in the left part of Fig. \ref{figure1}. Due to the matter component of molecular polaritons, exciton--exciton annihilation also plays a role in polaritonic systems and contributes to the overall rate of polariton annihilation.

In order to model exciton--exciton annihilation in polaritonic systems we extend the Tavis-Cummings model \cite{tavis_exact_1968, tavis_approximate_1969} by including intermolecular interactions and on-site disorder to obtain the polariton eigenstates, from which the annihilation rate can be calculated using Fermi's Golden rule \cite{tempelaar_excitonexciton_2017}. If the rotating wave approximation is employed when formulating the Hamiltonian, the system can be divided into subspaces with different numbers of excitations distributed between the excitonic and photonic subsystems. In this case, the exciton--exciton annihilation rate may be calculated from the two-particle eigenstates, which we denote by $|\Psi_{\alpha}\rangle$. Microscopically, the annihilation rate can then be described according to Fermi's Golden rule ($\hbar=1$)

\begin{equation}\label{rate_equation}
    \Gamma = \frac{2\pi}{\tau_\mathrm{nr}}\sum_{\alpha}\frac{1}{Z}e^{-\frac{\omega_{\alpha}}{k_BT}}\sum_{i=1}^N|\langle f_i|H_A|\Psi_{\alpha}\rangle|^2.
\end{equation}

The expression is a weighted sum of annihilation rates for every two-particle state $\frac{2\pi}{\tau_\mathrm{nr}}\sum_{i=1}^N|\langle f_i|H_A|\Psi_{\alpha}\rangle|^2$, where $|f_i\rangle$ models an energetically higher lying localized excited state of the $i$-th molecule, which are coupled to the two-particle eigenstates by an annihilation operator $H_A$. The annihilation operator $H_A$ represents a perturbation of the system, which may take different forms depending on the type of annihilation process. The density of states which is normally part of Fermi's Golden rule is replaced by the internal conversion rate between the molecular $|f\rangle$ and $|e\rangle$ states, $\frac{1}{\tau_\mathrm{nr}}$ \cite{ryzhov_low-temperature_2001, tempelaar_excitonexciton_2017}. The per state annihilation rates are weighted by Boltzmann factors $\frac{1}{Z}e^{-\frac{\omega_{\alpha}}{k_BT}}$ with the resonance frequency of the two-particle eigenstates $\omega_{\alpha}$ and the partition function $Z=\sum_{\alpha}e^{-\frac{\omega_{\alpha}}{k_BT}}$.

For the case of exciton--exciton annihilation, the annihilation Hamiltonian can be expressed as

\begin{equation}
    H_\text{ex-ex} = K \sum_{\langle i,j\rangle} \left(\tau_i^+\sigma_i^-\sigma_j^- + \tau_i^-\sigma_i^+\sigma_j^+\right),
\end{equation}

with the intermolecular annihilation coupling constant $K$, the lowering (raising) operator of the first excited state $|e_i\rangle$ of the \textit{i}-th molecule $\sigma_i^-$  ($\sigma_i^+$), and the lowering (raising) operator of the energetically higher lying excited state $|f_i\rangle$ of the \textit{i}-th molecule $\tau_i^-$  ($\tau_i^+$). We assume that exciton--exciton annihilation only occurs between nearest neighboring molecules, indicated by the sum notation $\langle i,j\rangle$ and that the intermolecular annihilation coupling is of dipolar form such that the corresponding coupling constant is given by $K = \frac{\tilde{\mu}_{eg}\tilde{\mu}_{fe}}{r^3}\kappa$, with the magnitude of the transition dipole moment of the transition from the ground to the first-excited state $\tilde{\mu}_{eg}$, the magnitude of the transition dipole moment of the $|e\rangle$ to $|f\rangle$ transition $\tilde{\mu}_{fe}$, the center of mass distance of neighboring molecules $r$, and an orientational factor $\kappa$, which determines the sign of the intermolecular couplings \cite{kasha_exciton_1965}. For the calculation of the annihilation rate, the sign of the coupling constant $K$ does not matter, since it enters the rate expression quadratically. Thus, we set $\kappa$ to $\kappa=1$. Furthermore, we define effective transition dipole moments by absorbing the distance dependence of the coupling into the transition dipole moments $\mu_{fi}=\frac{\tilde{\mu}_{fi}}{r^{\frac{3}{2}}}$ such that the intermolecular coupling constant takes the simple form $K = \mu_{eg}\mu_{fe}$.

Besides exciton--exciton annihilation, the additional coupling introduced by strong light--matter interactions imposes a second annihilation process, which couples the higher lying molecular $|f\rangle$ states to the electromagnetic field mode. In this process, a single excitation of the field mode and a singly excited molecule couple via light--matter interaction to form the molecular $|f\rangle$ state, which subsequently relaxes by internal conversion back to the $|e\rangle$ state (right part Fig. \ref{figure1}). We will call this process exciton--photon annihilation in the following. The corresponding annihilation Hamiltonian can be expressed as

\begin{equation}
    H_\text{ex-ph} = g_f \sum_{i=1}^N \left(\tau_i^+\sigma_i^-a + \tau_i^-\sigma_i^+a^\dagger\right),
\end{equation}

where $g_f$ is the corresponding light--matter annihilation coupling constant and $a$ ($a^\dag$) is the annihilation (creation) operator of the bosonic field mode. Within the dipole and long-wavelength approximations and if the polarization of the electric field is parallel to the transition dipole moments of the molecules, the light--matter coupling constant is given by 

\begin{equation}\label{xi}
    g_f = \sqrt{\frac{\omega_c}{2\epsilon_0V}}\tilde{\mu}_{fe} = \epsilon\mu_{fe}.
\end{equation}

Here, $\omega_c$ is the resonance frequency of the cavity, $\epsilon_0$ is the vacuum permittivity, and $V$ is the quantized mode volume of the electromagnetic field. For the second equality we define an effective vacuum electric field strength $\epsilon:=\sqrt{\frac{\omega_c}{2\epsilon_0V}}r^{\frac{3}{2}}$. Using our definitions, the effective transition dipole moments and the effective vacuum electric field strength have the same dimensions, making them directly comparable. Both annihilation processes, exciton--exciton and exciton--photon annihilation, induce annihilation rates, $\Gamma_\text{ex-ex}$ and $\Gamma_\text{ex-ph}$, respectively.

As mentioned above, we calculate the annihilation rates from the two-particle eigenstates of a generalized version of the Tavis-Cummings model, which describes the interaction between a linear chain of $N$ identical mutually coupled molecules and a single mode of an electromagnetic field. The system Hamiltonian without the perturbation of annihilation processes is given by ($\hbar=1$)

\begin{equation}\label{Hamiltonian}
    H_0 = H_m + \omega_ca^\dag a + g \sum_{i=1}^N\left(a^\dag\sigma_i^- + \sigma_i^+a\right),
\end{equation}

with the molecular Hamiltonian 

\begin{equation}\label{molecular_H}
H_m = \sum_{i=1}^N \left(\omega_m + \delta_i\right)\sigma_i^+\sigma_i^- +  J\sum_{\langle i,j\rangle} 
    \sigma_i^+\sigma_j^-,
\end{equation}


the light--matter coupling constant $g$, the intermolecular coupling constant $J$, the transition frequency of the ground to first excited state transition $\omega_m$, and the on-site energy disorder $\delta_i$ sampled from a uniform distribution on $[-\Omega, \Omega]$. The detuning will be defined as $\delta = \omega_c-\omega_\mathrm{exc}$, where $\omega_\mathrm{exc}$ is the resonance energy of the lowest eigenstate of the molecular Hamiltonian.
The second term in Eq. (\ref{molecular_H}) describes intermolecular couplings that induce the exciton hopping between the \textit{i}-th and \textit{j}-th singly excited molecule. We again assume that the coupling occurs only between nearest-neighboring molecules, indicated by the sum notation $\langle i,j\rangle$. Furthermore, we assume the same dipolar form for the intermolecular coupling as before, but set $\kappa=-1$ such that the corresponding coupling constant is given by $J = -\mu_{eg}^2$.

Here, the sign of the intermolecular coupling has a profound influence on the eigenstates of the system, since it determines whether the energetically lowest eigenstate of the molecular Hamiltonian has a large transition dipole moment or none. When additionally including light--matter interactions, the sign of the intermolecular coupling determines if the energetically lowest eigenstate has large photonic character regardless of the strength of the light--matter coupling. Here, we consider a negative intermolecular coupling strength characteristic of a J-type molecular aggregate \cite{valkunas_molecular_2013}. In this case, the energetically lowest eigenstate is always polaritonic in nature. 

The third term in Eq. (\ref{Hamiltonian}) corresponds to the light--matter interaction between the molecular excited states and the electromagnetic field. Similarly to Eq. (\ref{xi}) the coupling constant simplifies to

\begin{equation}\label{lm_coupling}
    g_e = \sqrt{\frac{\omega_c}{2\epsilon_0V}}\tilde{\mu}_{eg} = \epsilon\mu_{eg}.
\end{equation}

For the Tavis-Cummings model, it is convenient to introduce the collective coupling strength $g_{e,c}:=g_e\sqrt{N}$, which represents half of the energetic splitting of the polaritonic states.

In order to calculate the annihilation rates, the Hamiltonian matrix of the generalized Tavis-Cummings model is diagonalized within the two-particle manifold to obtain the eigenstates and energies. Generally, an eigenstate within the two-particle manifold is expanded as

\begin{equation}
    |\Psi_{\alpha}\rangle = c^{(\alpha)}_0 |2,0\rangle+\sum_{i=1}^Nc^{(\alpha)}_i|1,e_i\rangle + \sum_{i=1}^N\sum_{j>i}^Nc^{(\alpha)}_{i,j}|0,e_ie_j\rangle,
\end{equation}

where $|n,e_i\cdots e_k\rangle$ are basis state with $n$ quanta in the cavity field mode and the \textit{i}-th to \textit{k}-th molecule excited and $c^{(\alpha)}_0$, $c^{(\alpha)}_i$, and $c^{(\alpha)}_{i,j}$ are the corresponding expansion coefficients. Based on the basis expansion of the eigenstates the matrix elements of the annihilation operators can be simplified. For matrix elements of the exciton--exciton annihilation operator this yields

\begin{equation}\label{ex-ex}
    \sum_{i=1}^N|\langle f_i|H_\text{ex-ex}|\Psi_{\alpha}\rangle|^2 = K^2\sum_{i=1}^{N}|c^{(\alpha)}_{i,i+1}+c^{(\alpha)}_{i-1,i}|^2,
\end{equation}

where we define $c^{(\alpha)}_{0,1}=c^{(\alpha)}_{N,N+1}=0$. Thus, the exciton--exciton annihilation rate is given by

\begin{equation}
    \Gamma_\text{ex-ex} = \frac{2\pi}{\tau_\mathrm{nr}}\frac{K^2}{Z}\sum_{\alpha}e^{-\frac{\omega_{\alpha}}{k_BT}}\sum_{i=1}^{N}|c^{(\alpha)}_{i,i+1}+c^{(\alpha)}_{i-1,i}|^2.
\end{equation}

Since exciton--exciton annihilation necessitates the interaction between two molecules, only the expansion coefficients of the purely matter basis states enter into the corresponding matrix elements. Similarly, the matrix elements of the exciton--photon annihilation operator simplify to

\begin{equation}\label{ex-ph}
    \sum_{i=1}^N|\langle f_i|H_\text{ex-ph}|\Psi_{\alpha}\rangle|^2 = g_f^2\sum_{i=1}^N|c^{(\alpha)}_{i}|^2
\end{equation}

such that the corresponding rate is given by

\begin{equation}
    \Gamma_\text{ex-ph} = \frac{2\pi}{\tau_\mathrm{nr}}\frac{g_f^2}{Z}\sum_{\alpha}e^{-\frac{\omega_{\alpha}}{k_BT}}\sum_{i=1}^{N}|c^{(\alpha)}_{i}|^2.
\end{equation}

Here, molecules and photons need to interact to enable annihilation, thus the corresponding matrix elements are proportional to the contribution of mixed photonic--molecular basis states. Based on these equations in conjunction with Eq. (\ref{rate_equation}) exciton--exciton and exciton--photon annihilation rates may be calculated from the two-particle states of the generalized Tavis-Cummings model. 

\begin{figure}[h!]
  \includegraphics[]{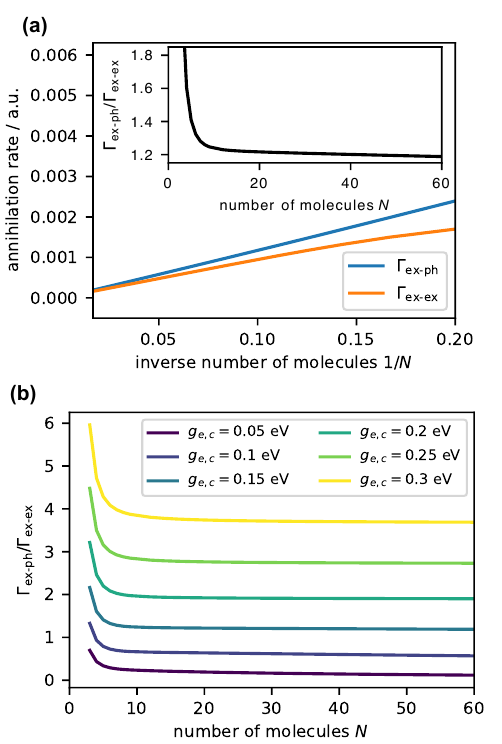}
  \caption{\label{figure2} Exciton--exciton and exciton--photon annihilation rates as functions of molecular number and collective light–matter coupling strength. (a) Exciton--exciton $\Gamma_\text{ex-ex}$ and exciton--photon $\Gamma_\text{ex-ph}$ annihilation rates depending on the inverse number of coupled molecules for typical parameters of a system in the strong light--matter coupling regime. The inset shows the dependence of the ratio of both annihilation rates on the number of coupled molecules. (b) Ratio of exciton--photon and exciton--exciton annihilation rates depending on the number of coupled molecules for various values of the collective light--matter coupling strength. For a given number of molecules $N$, the resonance energy of the cavity is always tuned to the first eigenstate of the molecular Hamiltonian $H_m$.}
\end{figure}

In Fig. \ref{figure2}(a) we show the dependence of exciton--exciton and exciton--photon annihilation rates on the inverse number of molecules for the extended Tavis-Cummings model without on-site disorder, i.e., $\Omega=0$. To obtain this plot, we set $J=-K=-\mu_{eg}^2=-0.1\ \mathrm{eV}$, $g_{e,c} = 0.15\ \mathrm{eV}$ and $g_e = g_f = g_{e,c}/\sqrt{N}$ to ensure that the collective coupling strength $g_{e,c}$ is unchanged while increasing the number of molecules. Furthermore, we set $\omega_m=2\ \mathrm{eV}$ and set the resonance frequency of the electromagnetic field mode $\omega_c$ to the lowest eigenvalue of the molecular Hamiltonian $H_m$ such that there is no detuning to the state with the largest transition dipole moment. Lastly, we take the thermal distribution $\frac{1}{Z}e^{-\frac{\omega_{\alpha}}{k_BT}}$ for a temperature of 293 K. 

Both annihilation rates, $\Gamma_\text{ex-ex}$ and $\Gamma_\text{ex-ph}$, scale linearly with $1/N$. For $\Gamma_\text{ex-ex}$, this can be rationalized by considering that there are $\binom{N}{2}\sim N^2$ basis states with two distinct excited molecules. Out of these basis states only those where two neighboring molecules are excited contribute to the exciton--exciton annihilation rate because we consider $H_\text{ex-ex}$ to only couple nearest neighbors. Since there are only $N-1\sim N$ basis states where two neighboring molecules are excited, but $\binom{N}{2}\sim N^2$ two-particle basis states overall, the annihilation rate scales with $1/N$ if the two-particle eigenstates are almost uniformly delocalized. For $\Gamma_\text{ex-ph}$ the $1/N$ scaling can be rationalized by considering that the coupling constant $g_f$ enters quadratically into the annihilation rate and decreases as $1/\sqrt{N}$, if the collective coupling strength is kept constant. Therefore, the exciton--photon annihilation rate also decreases as $1/N$. Since both annihilation rates show the same asymptotic behavior with regard to the number of molecules, their ratio converges to a constant, as shown in the inset of Fig. \ref{figure2}(a). Since in any experiment the number of coupled molecules is really large, we assume that the limit of the ratio corresponds to the experimental situation. As the ratio converges to a value different from zero, exciton--exciton as well as exciton--photon annihilation contribute to the multi-particle dynamics in polaritonic systems and which of both is the dominant contribution can be directly inferred from the limit of the ratio.

From Fig. \ref{figure2}(b) it is evident that the ratio of the annihilation rates converges also for a range of different collective light--matter coupling strengths. For small collective coupling strengths $g_{e,c}$, exciton--exciton annihilation dominates over exciton--photon annihilation, i.e., $\Gamma_\text{ex-ph}/\Gamma_\text{ex-ex}<1$. By increasing the collective light--matter coupling, the ratio $\Gamma_\text{ex-ph}/\Gamma_\text{ex-ex}$ increases and consequently, exciton--photon annihilation becomes more prominent. At a critical value of the collective light--matter coupling the ratio switches from $\Gamma_\text{ex-ph}/\Gamma_\text{ex-ex}<1$ to $\Gamma_\text{ex-ph}/\Gamma_\text{ex-ex}>1$ [Fig. \ref{figure2}(b)]. Thus, exciton--photon annihilation becomes the predominant annihilation process for large collective light--matter coupling strengths. In the case of $g_c=0.3\ \mathrm{eV}$, the rate of exciton--photon annihilation is four times larger than that of exciton--exciton annihilation [Fig. \ref{figure2}(b)]. However, the increase in the ratio of annihilation rates is smaller than expected by the quadratic $g_f$ scaling of $\Gamma_\text{ex-ph}$. If the collective light--matter coupling strength is increased, the molecular component of the wavefunctions become more uniformly delocalized compared to the wavefunctions of the purely molecular Hamiltonian $H_m$. This is clear since the molecular component of the polaritonic eigenstates in the Tavis-Cummings model without intermolecular interactions is perfectly delocalized, i.e., all expansion coefficients of the molecular basis states are equal. The enhanced delocalization leads to an increase in the exciton--exciton annihilation rate $\Gamma_\text{ex-ex}$ for increasing collective light--matter coupling and hence to the observed deviation from the expected quadratic scaling of the ratio of annihilation rates.

In the following, we investigate how the limit of the ratio of both annihilation rates depends on the system parameters and especially, at which parameter set the transition from exciton--exciton dominated to exciton--photon dominated annihilation occurs, i.e., $\Gamma_\text{ex-ph}/\Gamma_\text{ex-ex}=1$. Hereby, we also include on-site disorder by setting $\Omega=0.1\ \mathrm{eV}$ and average every calculation over 100 disorder realizations to obtain mostly converged results while keeping the computational demand low. To analyze the dependence of the annihilation rates on the system parameters, it is mandatory to find a minimal set of independent molecular and photonic parameters, which still fully determines the couplings $J$, $K$, $g_e$, and $g_f$. First of all, we notice that $g_f$ and $K$ both depend linearly on the magnitude of the transition dipole moment from the first to the higher excited molecular state, $\mu_{fe}$. Thus, the ratio of annihilation rates is actually independent of $\mu_{fe}$. Similarly, the factor $\frac{2\pi}{\tau_\mathrm{nr}}$ cancels by forming the ratio of the annihilation rates. Furthermore, according to our considerations above, all coupling parameters solely depend on the magnitude of the effective transition dipole moment from the ground to first excited molecular state $\mu_{ge}$ and the effective vacuum electric field strength of the cavity mode $\epsilon$. This leaves us with three independent parameters, the effective transition dipole moment $\mu_{eg}$, the effective vacuum electric field strength $\epsilon$, and the resonance energy of the cavity $\omega_c$, to fully determine the ratio of exciton--photon and exciton--exciton annihilation rates. 

\begin{figure}[h!]
  \includegraphics[]{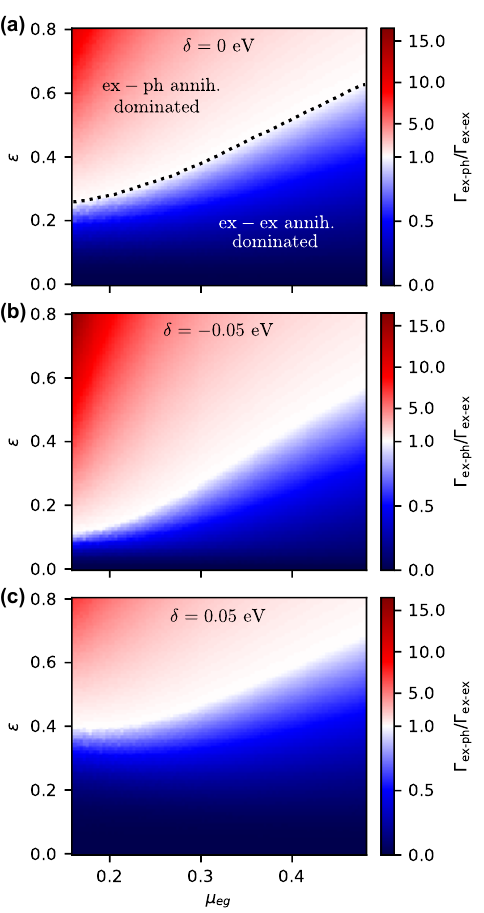}
  \caption{\label{figure3} Ratio of exciton--photon and exciton--exciton annihilation rates depending on the system parameters, i.e., the magnitude of the effective transition dipole moment from the ground to the first excited state $\mu_{eg}$ of the molecules and the effective vacuum electric field strength of the quantized field $\epsilon$. Three cases are considered, where the resonance energy of the cavity is (a) tuned to the first eigenstate of the molecular Hamiltonian $H_m$, (b) negatively detuned, or (c) positively detuned.}
\end{figure}

\begin{figure}[h!]
  \includegraphics[]{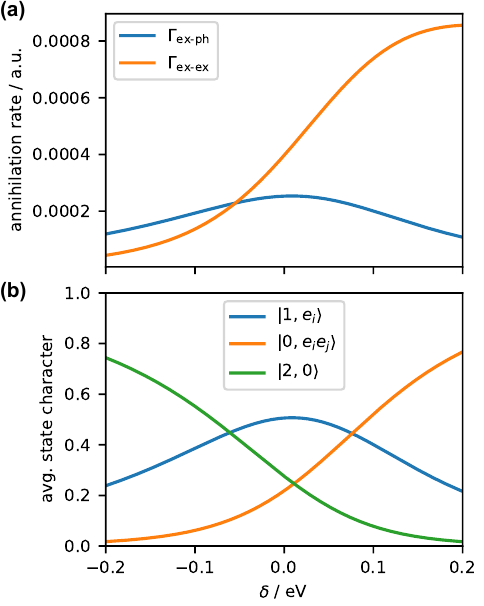}
  \caption{\label{figure4} Effect of detuning between the cavity and the molecular subsystem. (a) Dependence of the exciton--photon and exciton--exciton annihilation rates on the detuning between the cavity and the molecular subsystem with system parameters $\mu_{eg}=\sqrt{0.1\ \mathrm{eV}}$ and $\epsilon=\sqrt{0.1\ \mathrm{eV}}$. (b) Dependence of the Boltzmann averaged state character on the detuning between the cavity and the molecular subsystem with the same system parameters as in (a). The Boltzmann averaged state character is defined by $P_I = \sum_\alpha \frac{1}{Z}e^{-\frac{\omega_{\alpha}}{k_BT}} \sum_{i\in I} |c^{(\alpha)}_i|^2$, where $I$ is an index set indexing either purely photonic ($|2,0\rangle$), exciton--photon ($|1,e_i\rangle$), or purely molecular ($|0,e_ie_j\rangle$) basis states.}
\end{figure}

Thus, we calculated the ratio of exciton--photon and exciton--exciton annihilation rates depending on the effective transition dipole moment and the effective vacuum electric field strength for three different values of the detuning between the cavity resonance energy and the energy of the first excited state of the molecular subsystem. In Fig. \ref{figure3}(a) we show the dependence of the ratio of exciton--photon and exciton--exciton annihilation rates on $\mu_{eg}$ and $\epsilon$, where we assume that the resonance energy of the cavity is in resonance with the energetically lowest eigenstate of the molecular Hamiltonian $H_m$. We set the number of molecules to $N=20$ to keep the computational demand low while being already close to the saturation limit as evident from Fig. \ref{figure2}(b). The results in Fig. \ref{figure3}(a) demonstrate that there are generally two different annihilation regimes: while in the blue area, exciton--exciton annihilation is the dominant annihilation process, in the red area excitons and photons recombine predominantly. As expected, exciton--exciton annihilation is dominant if $\mu_{eg} \gg \epsilon$ and conversely exciton--photon annihilation is dominant if $\mu_{eg} \ll \epsilon$. The transition from exciton--exciton to exciton--photon dominated annihilation occurs always at a field strength slightly higher than the corresponding transition dipole moment. 

In Fig. \ref{figure3}(b) and \ref{figure3}(c) we show that the transition regime between exciton--exciton and exciton--photon dominated annihilation strongly depends on the detuning between the resonance energy of the cavity and the energetically lowest eigenstate of the molecular Hamiltonian. In Fig. \ref{figure3}(b) the resonance energy of the cavity is slightly negatively detuned from the energetically lowest excitonic transition. Consequently, exciton--photon annihilation begins to dominate already at lower vacuum electric field strengths for a given transition dipole moment compared to the resonant case. However, if the resonance of the cavity is slightly positively detuned from the energetically lowest excitonic transition, as in Fig. \ref{figure3}(c), exciton--photon annihilation only dominates for larger vacuum electric field strengths compared to the resonant case. This behavior reflects the dependence of exciton--photon and exciton--exciton annihilation on the detuning between the cavity and the molecular subsystem as shown in Fig. \ref{figure4} together with the Boltzmann averaged state character. Starting from a negatively detuned system, an increase of the detuning leads to an increase of the average purely molecular state character [Fig. \ref{figure4}(b)] and thus, the exciton--exciton annihilation rate increases strictly with an increasing detuning [Fig. \ref{figure4}(a)]. Compared to that, the exciton--photon annihilation rate first increases with increasing detuning until the cavity and the molecular subsystem are in resonance, i.e., $\delta=0$, where the annihilation rate takes on its maximum [Fig. \ref{figure4}(a)]. If the detuning is further increased, the annihilation rate decreases again. The exciton--photon annihilation rate closely follows the average exciton--photon state character, which is maximal if the cavity is in resonance with the molecular excitonic state with the largest transition dipole moment [Fig. \ref{figure4}(b)].

Interestingly, the shift of the transition regime induced by variation of the detuning is larger for small transition dipole moments. This effect occurs since an increase of the transition dipole moment does not only lead to stronger intermolecular interaction but also to an increase of the light--matter coupling. Subsequently, larger light--matter coupling results in slower variations of the wave function with detuning.

In summary, we have derived a microscopic theory for the annihilation of polaritons and investigated the dependence of the annihilation rate on various system parameters. While in purely molecular systems, exciton--exciton annihilation solely determines the annihilation dynamics, we show that strong light--matter coupling leads to the emergence of an additional annihilation process, which we call exciton--photon annihilation. Therefore, the annihilation of polaritons is determined by a competition between exciton--exciton and exciton--photon annihilation. While exciton--exciton annihilation cannot be neglected, especially for large vacuum electric field strengths, exciton--photon annihilation is the predominant annihilation process in polaritonic systems. In a recent experimental study, the annihilation rates of cavity samples containing LH2 complexes in different light--matter coupling regimes were compared with reference samples without a cavity on glass substrates \cite{wu_efficient_2025}. The cavity sample in the strong-coupling regime showed a much larger annihilation rate compared to that of the reference sample and even in the weak coupling regime an increased annihilation rate has been reported compared to the reference sample. Our explanation for increased annihilation in such systems by the occurrence of exciton--photon annihilation provides a theoretical basis for the description of the multi-particle dynamics, which was missing so far.

In another recent experimental study, we also investigated the annihilation dynamics in a polaritonic system \cite{buttner_probing_2025}. However, we did not observe a significant difference of the annihilation rates in the polaritonic system and a reference sample without light--matter interactions. We explained this unexpected observation by a fast relaxation of the polaritonic states into a manifold of dark states (DS), which inevitably arise when an ensemble of molecules is coupled to a strong electromagnetic field \cite{gonzalez-ballestero_uncoupled_2016, botzung_dark_2020, scholes_entropy_2020, chen_tracking_2025, virgili_ultrafast_2011, xiang_state-selective_2019, georgiou_generation_2018}. Since DSs are purely molecular, they are not affected by exciton--photon annihilation. Therefore, only exciton--exciton annihilation occurs within the DS manifold, explaining why no change in the annihilation rate is observed in spite of light--matter coupling. 

In general, annihilation in excitonic systems can be used as a probe to quantify energy transport due to the locality of exciton--exciton interactions, since excitons have to diffuse and meet to annihilate. While the occurrence of exciton--exciton annihilation within polaritonic systems still assures a correlation between annihilation and transport, the connection is complicated by the occurrence of exciton--photon annihilation and its non-local nature, i.e., all molecules uniformly couple with similar coupling strength to the electromagnetic field. However, we think that increased annihilation rates through the occurrence of exciton--photon annihilation reveal the hybrid light--matter character of polaritons. In this way, exciton--photon annihilation represents an experimental fingerprint of polaritons and might be utilized to reveal whether one really observes polaritons in an experiment or some  untargeted effect as for example the response of DSs.

\begin{acknowledgments}
 L.N.P. acknowledges a fellowship by the \textit{Fonds der Chemischen Industrie} (FCI).
\end{acknowledgments}

\bibliography{Polariton_Annihilation}

\end{document}